\documentclass[onclumn]{revtex4}

\usepackage{graphicx}
\usepackage{amsmath}
\usepackage{bm}
\usepackage{multirow}
\usepackage{comment}

\def\title#1{\begin{center}{\Large\bf #1}\end{center}}
\def\author#1{\vskip 5mm \begin{center}{#1}\end{center}}
\def\address#1{\begin{center}{\it #1}\end{center}}
\newcommand{\simgt}{\lower.5ex\hbox{$\; \buildrel > \over \sim \;$}}
\newcommand{\simlt}{\lower.5ex\hbox{$\; \buildrel < \over \sim \;$}}

\begin{document}

\title{Quantum Larmor radiation in conformally flat universe}

\author{Rampei Kimura, Gen Nakamura, Kazuhiro Yamamoto}
\email[email:]{kazuhiro@hiroshima-u.ac.jp}

\address{
Department of Physical Science, Hiroshima University,
Higashi-Hiroshima 739-8526,~Japan}

\begin{abstract}
We investigate the quantum effect on the Larmor radiation from a moving 
charge in an expanding universe based on the framework of 
the scalar quantum electrodynamics (SQED). 
A theoretical formula for 
the radiation energy is derived at the lowest order of the 
perturbation theory with respect to the coupling constant of the SQED. 
We evaluate the radiation energy on the background universe so that 
the Minkowski spacetime transits to the Milne universe, 
in which the equation of motion for the mode function of the free 
complex scalar field can be exactly solved in an analytic way. Then,
the result is compared with the WKB approach, in which the equation 
of motion of the mode function is constructed with the WKB approximation
which is valid as long as the Compton wavelength is shorter than the 
Hubble horizon length.  
This demonstrates that the quantum effect on the Larmor radiation of the order
$e^2\hbar$ is determined by a non-local integration in time depending on
the background expansion. 
We also compare our result with a recent work
by Higuchi and Walker [Phys.~Rev.~D{\bf 80} 105019 (2009)], 
which investigated the quantum correction to the Larmor 
radiation from a charged particle in a non-relativistic motion 
in a homogeneous electric field.

\end{abstract}

\maketitle


\def\bfp{{\bf p}}
\def\bfk{{\bf k}}
\def\bfx{{\bf x}}
\def\bfV{{\bf V}}
\def\bfz{{z}}
\def\pxi{{\xi}}
\def\barm{{\bar m}}
\def\bark{{\kappa}}
\section{Introduction}
Recently, the quantum electrodynamics (QED) processes in a strong field 
background attract interest of theoretical researchers 
as well as experimental researchers \cite{Iso,Homma,Gies}. 
Especially, recent developments of strong laser technique might 
provide us 
with a chance to test the Schwinger effect, which causes a pair 
creation phenomenon in the strong laser-electric field \cite{Schwinger}. 
On the other hand, some authors claimed that the Unruh effect 
\cite{Unruh,HiguchiPR} might be detected by measuring the radiation 
from electron beam in a strong laser background 
\cite{ChenTajima,Schutzhold}, but it seems to be necessary for 
more careful investigations for the definite conclusion of the
relevance \cite{YNK}. 

An analogy between the 
radiation process from a moving charge in an expanding universe
and the classical Larmor radiation in the electro-magnetic 
theory has been investigated (\cite{NSY}, see also \cite{Higuchicfu,futamase}). 
The motion of a massive charged particle in an expanding background spacetime
can be regarded as an accelerated motion because the physical momentum of 
the particle changes as the background spacetime expands or contracts. 
This leads to the Larmor radiation from the charge in an accelerated motion
in the classical electro-magnetic theory. 
In Ref.~\cite{NSY}, it was also demonstrated that 
the quantum effect affects the radiation energy using a model 
on a background universe with specific expansion, where the equation of motion 
of the mode function is solved in an analytic manner exactly. 

In this paper, we focus our investigation on the quantum 
effect on the Larmor radiation from a moving charge in an 
expanding universe. Recently, Higuchi and Walker investigated
the first order quantum effect of the Larmor radiation from a 
moving charge in a homogeneous electric field background \cite{HW}. 
In Ref.~\cite{HW}, 
the motion of the moving charge is limited to the non-relativistic 
case. A generalization of the work of \cite{HW} is presented in 
Ref.~\cite{YNK}. In the present paper, we apply the method developed 
in Ref.~\cite{YNK} to the radiation from a moving charge 
in an expanding universe.

This paper is organized as follows: In Sec.~2, 
we derive a general formula for the radiation energy of the 
quantum Larmor process in an expanding universe on the basis 
of the scalar QED.
This formula, which is evaluated at the lowest order of the 
perturbation theory with respect to the coupling constant, 
is obtained by solving the equation of motion of the mode function
of the complex scalar field in an expanding universe. 
In Sec.~3, we consider the complex scalar field on the background 
universe that the Minkowski spacetime transits to the Milne universe,
on which the equation of motion of the mode function is exactly solved in an 
analytic manner. Then, the radiation energy is evaluated by performing
an integration numerically.
In Sec.~4, we adopt the WKB approach for the mode function, which
is constructed in an expanded form with respect to $\hbar$.
This approach yields the formula of the radiation energy in an expanded 
form with respect to $\hbar$ as well.
The quantum effect of the order of $\hbar$ is determined 
by the non-local nature in time of the background expansion. 
The result is compared with the work \cite{HW}.
Section 5 is devoted to the summary and conclusions.
In Appendix A, we summarized useful formulas in our derivation of 
the radiation formula.
Through out this paper, we set the speed of light $c=1$ and the 
metric convention $(+,-,-,-)$.

\section {formulation}
First, we derive the formula for the radiation energy
from a moving charge in the spatially flat 
Friedmann-Robertson-Walker spacetime, whose line 
element is given by 
\begin{eqnarray}
ds^2=a^2(\eta)[d\eta^2-\delta_{ij}dx^idx^j],
\end{eqnarray}
where $\eta$ is the conformal time and $a(\eta)$ is the scale factor. 
We consider the scalar QED action conformally coupled to the curvature
\begin{eqnarray}
S=\int d\eta d\bfx \sqrt{-g}\left[g^{\mu\nu}\left(\nabla_\mu-{ieA_\mu\over \hbar}
\right)\phi^\dagger\left(\nabla_\nu+{ieA_\nu\over \hbar}
\right)\phi-\left({m^2\over \hbar^2}-\xi R\right)\phi^*\phi-{1\over 4\mu_0}
F_{\mu\nu}F^{\mu\nu}
\right],
\label{actionb}
\end{eqnarray}
where $F_{\mu\nu}$ is the field strength, $\mu_0$ is the magnetic permeability
of vacuum, and $\xi=1/6$. Introducing the conformally rescaled field $\Phi$,
\begin{eqnarray}
\phi={\Phi\over a(\eta)},
\end{eqnarray}
we may rewrite the action as 
\begin{eqnarray}
S=\int d\eta d\bfx \left[
\left(D_\mu \Phi\right)^\dagger D^\mu \Phi
-{m^2a^2(\eta)\over \hbar^2 }\Phi^\dagger\Phi-{1\over 4}F_{\mu\nu}F^{\mu\nu}
\right],
\label{actiona}
\end{eqnarray}
where $D_\mu\equiv \partial_\mu+{ie}A_\mu/{\hbar}$. 
It might be worthwhile to note  that the equivalence between 
Eqs.~(\ref{actionb}) and (\ref{actiona}) will break down at 
one-loop order \cite{Higuchicfu}. However, this does not
affect our result because our investigation is at tree level.

We adopt the in-in formalism \cite{Weinberg} to evaluate the 
radiation energy, in which the expectation value of the 
Heisenberg operator $Q$ at the time $\eta$ is given by the 
following form,
\begin{eqnarray}
 \left\langle Q(\eta) \right\rangle=\sum^\infty_{N=0}{i^N\over \hbar^N}
\int^\eta_{-\infty}
d\eta_N\ldots\int^{\eta_2}_{-\infty}d\eta_1\left\langle [H_I(\eta_1),\ldots[H_I(\eta_N),Q(\eta)]\ldots]\right\rangle,
\label{Q}
\end{eqnarray}
where $H_I$ denotes the interaction Hamiltonian operator, 
\begin{eqnarray}
H_I(\eta)&=&-\frac{ie}{\hbar}\int d^3{\bf x}A^\mu(\eta,\bfx)
[\partial_\mu\Phi^\dagger(\eta,\bfx)\Phi(\eta,\bfx)
-\Phi^\dagger(\eta,\bfx) \partial_\mu\Phi(\eta,\bfx)],
\label{HI}
\end{eqnarray}
and the operators in the right hand side of Eq.~(\ref{Q})
are those of the interaction picture. 
The free field of the electromagnetic field quantized in a 
finite box is 
\begin{eqnarray}
 A_\mu=\sqrt{\mu_0\hbar\over V}\sum_{\lambda=1,2}\sum_{\bfk}
\sqrt{\frac{1}{2k}}\epsilon^\lambda_\mu
\left(a^\lambda_\bfk e^{-ik\eta}+a^{\lambda\dagger}_{-\bfk} e^{ik\eta}\right)e^{i\bfk\cdot\bfx},
\end{eqnarray}
where $\epsilon^\lambda_\mu$ denotes the polarization tensor, 
$\lambda$ denotes the direction of polarization, $V$ is 
the volume of the box, $a^{\lambda\dagger}_\bfk$ and $a^{\lambda}_\bfk$ 
are the creation and annihilation operator, respectively, 
which satisfy the following commutation relation,
\begin{eqnarray}
 [a^\lambda_\bfk,a^{\dagger\lambda^\prime}_{\bfk^\prime}]=\delta_{\lambda\lambda^\prime}\delta_
{\bfk,\bfk^\prime}.
\end{eqnarray}
In the similar way, the quantized complex scalar field is given by
\begin{eqnarray}
\Phi(x)=\sum_{\bfp}\sqrt{\frac{\hbar}{V}}e^{i\bfp\cdot\bfx/\hbar}
\Bigl(\varphi_\bfp(\eta)b_\bfp+\varphi^*_\bfp(\eta) c^\dagger_{-\bfp}\Bigr),
\end{eqnarray}
where the mode function $\varphi_\bfp(\eta)$ obeys the equation of motion,
\begin{eqnarray}
&&\left({\partial^2 \over \partial \eta^2}
+{\bfp^2+m^2a^2(\eta)\over \hbar^2}\right)\varphi_\bfp(\eta)=0,
\label{eqmod}
\end{eqnarray}
with the normalization condition 
$\dot \varphi^*_\bfp\varphi_\bfp-\dot \varphi_\bfp\varphi_\bfp=i$,
and the creation and annihilation operators satisfy the commutation relation,
\begin{eqnarray}
 [b_\bfp,b^\dagger_{\bfp^\prime}]=\delta_{\bfp,\bfp^\prime},\quad
 [c_\bfp,c^\dagger_{\bfp^\prime}]=\delta_{\bfp,\bfp^\prime},\quad
[{\rm OTHER}]=0.
\end{eqnarray}
\begin{figure*}[t]
\includegraphics[scale=0.8]{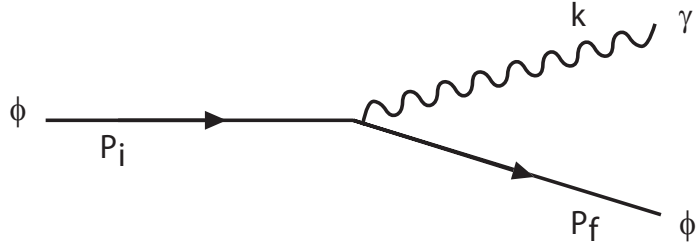}%
\caption{Feynman Diagram for the process.
\label{fig1} {}}
\end{figure*}

Choosing the Heisenberg operator $Q$ as the number operator 
$a^{\lambda\dagger}_\bfk a^{\lambda}_\bfk$, we calculate the total 
radiation energy as follows,
\begin{eqnarray}
E= \int d^3\bfk\hbar k\sum_{\lambda}
\langle a^{\lambda\dagger}_\bfk a^{\lambda}_\bfk\rangle
=-\int d^3\bfk\hbar k\sum_{\lambda}\hbar^{-2}\int^t_{-\infty}dt_2\int^{t_2}_{-\infty}
dt_1\left\langle \bigl[H_I(t_1),[H_I(t_2),
a^{\lambda\dagger I}_\bfk a^{\lambda I}_\bfk]\bigr]\right\rangle,
\label{Ea}
\end{eqnarray}
where we only considered the contribution up to the 
order of $e^2$. We set the initial state as,
\begin{eqnarray}
 |i\rangle=|0\rangle_{a}\otimes|1\rangle_{b}\otimes|0\rangle_{\bar{b}},
\end{eqnarray}
where the subscript $a$ means the photon number basis, $b$ 
is the charged particle number basis, and $\bar{b}$ is the 
anti-particle number basis.

We consider the lowest order contribution of the process so 
that one photon is emitted from the charged particle, as shown 
in Fig.~\ref{fig1}. From Eq.~(\ref{Ea}), we have
\begin{eqnarray}
E&=&- {e^2\over \epsilon_0} \int {d^3k\over (2\pi)^3}  k\biggl\{
\biggl|\int d\eta {e^{ik\eta}\over {\sqrt{2k}}} \Bigl(
{\partial\over\partial \eta}\varphi_{\bfp_f}(\eta)^* \varphi_{\bfp_i}(\eta)-
\varphi_{\bfp_f}^*(\eta){\partial\over\partial \eta}\varphi_{\bfp_i}(\eta)
\Bigr)\biggr|^2
\nonumber
\\
&&~~~~~~~~~~~~~~~~~~
-\biggl|\int d\eta {e^{ik\eta}\over {\sqrt{2k}}} \Bigl(
{i\bfp_f\over\hbar}\varphi_{\bfp_f}(\eta)^* \varphi_{\bfp_i}(\eta)+
\varphi_{\bfp_f}^*(\eta){i\bfp_i\over\hbar}\varphi_{\bfp_i}(\eta)
\Bigr)\biggr|^2
\biggr\},
\label{radiationenergy}
\end{eqnarray} 
where $\bfp_f=\bfp_i-\hbar\bfk$, $\hbar\bfk$ is the photon momentum, 
and $\epsilon_0$ is the permittivity
of vacuum, which is related to $\mu_0$ by $\epsilon_0\mu_0=1/c^2=1$. 
Using the equation of motion (\ref{eqmod}), Eq.~(\ref{radiationenergy})
leads to
\begin{eqnarray}
E&=&- {e^2\over \epsilon_0} \int {d^3k\over (2\pi)^3}   k\biggl\{
\biggl|\int d\eta {e^{ik\eta}\over\sqrt{2k}} {{\hat \bfk}\cdot(\bfp_i+\bfp_f)
\over \hbar}\varphi_{\bfp_f}^*(\eta)\varphi_{\bfp_i}(\eta)
\biggr|^2
\nonumber
\\
&&~~~~~~~~~~~~~~~~~~-\biggl|\int d\eta  {e^{ik\eta}\over\sqrt{2k}}
{\bfp_i+\bfp_f
\over \hbar}
\varphi_{\bfp_f}^*(\eta)\varphi_{\bfp_i}(\eta)
\biggr|^2
\biggr\},
\label{radiationenergy2}
\end{eqnarray}
where $\hat \bfk=\bfk/|\bfk|$.
Furthermore, Eq.~(\ref{radiationenergy2}) reduces to (cf. \cite{NSY})
\begin{eqnarray}
E&=& {e^2\over 2\epsilon_0} \int {d^3k\over (2\pi)^3}   
{4(\bfp_i^2-(\bfp_i\cdot\hat\bfk)^2)\over \hbar^2}
\biggl|\int d\eta {e^{ik\eta}} \varphi_{\bfp_f}^*(\eta)\varphi_{\bfp_i}(\eta)
\biggr|^2.
\label{radiationenergy3}
\end{eqnarray}
Note that this is the energy in the conformally 
rescaled spacetime, 
which is not the physical energy. We 
can read the physical radiation energy as ${\cal E} = E/a$. 
\section{Solvable model in Minkowski-Milne universe}
In this section, we consider the universe whose scale factor 
is written in the form, 
\begin{eqnarray}
a(\eta)=\sqrt{1+e^{2\rho \eta}},
\label{amilne}
\end{eqnarray}
where $\rho$ is the parameter to describe the expansion rate. 
This gives the Minkowski spacetime at $\eta\rightarrow-\infty$,
and the Milne universe at $\eta\rightarrow+\infty$.
The equation of motion of the mode function (\ref{eqmod}) is 
\begin{eqnarray}
\left[{d^2\over d\eta^2}+{p^2+m^2(1+e^{2\rho\eta})\over \hbar^2}\right]
\varphi_p(\eta)=0.
\end{eqnarray}
The solution is given by using the Bessel function,
\begin{eqnarray}
\varphi_p(\eta)=\sqrt{\pi\over 2\rho \sinh\pi\pxi}
J_{-i\pxi}\left(\barm e^{\rho \eta}\right),
\end{eqnarray}
where 
\begin{eqnarray}
\pxi=\sqrt{\bfp^2+m^2\over \rho^2\hbar^2}, ~~~~~~~\barm={m\over \rho\hbar}.
\nonumber
\end{eqnarray}

The radiation energy is given by the formula (\ref{radiationenergy3}),
then we focus on 
\begin{eqnarray}
\int_{-\infty}^\infty d\eta e^{ik\eta}\varphi_{\bfp_f}^*(\eta)
\varphi_{\bfp_i}(\eta)
&=&\sqrt{\pi\over 2\rho\sinh \pi\xi_f}\sqrt{\pi\over 2\rho\sinh \pi\xi_i}
\int_{-\infty}^\infty d\eta e^{ik\eta}
J_{i\xi_f}(\barm e^{\rho\eta})J_{-i\xi_i}(\barm e^{\rho\eta}),
\nonumber
\\
&=&\sqrt{\pi\over 2\rho\sinh \pi\xi_f}\sqrt{\pi\over 2\rho\sinh \pi\xi_i}
{\barm^{-ik/\rho}\over\rho}
\int_0^\infty dz z^{ik/\rho-1}
J_{i\xi_f}(z)J_{-i\xi_i}(z),
\end{eqnarray}
where we defined
\begin{eqnarray}
\xi_i=\sqrt{\bfp^2_i+m^2\over \rho^2\hbar^2}, ~~~~~~
\xi_f=\sqrt{\bfp^2_f+m^2\over \rho^2\hbar^2}=\sqrt{\bfp_i^2-2\hbar kp_i\cos\theta+\hbar^2 k^2+m^2\over \rho^2\hbar^2}.
\end{eqnarray}
We use the mathematical formula \cite{Gradshteyn}
\begin{eqnarray}
\int_0^\infty dz z^{ik/\rho-1}
J_{i\xi_f}(z)J_{-i\xi_i}(z)=
{2^{i\bark-1}\Gamma(1-i\bark)\Gamma(i(\xi_f-\xi_i+\bark)/2)
\over 
\Gamma(i(\xi_f-\xi_i-\bark)/2+1)\Gamma(i(\xi_f+\xi_i-\bark)/2+1)
\Gamma(i(-\xi_f+\xi_i-\bark)/2+1)},
\end{eqnarray}
where we defined $\bark=k/\rho$.
Then, the radiation energy is
\begin{eqnarray}
E&=&{e^2 p_i^2\over 2\pi\epsilon_0 \rho^4}\int_0^\infty dk k^2 
\int_{-1}^1d\cos\theta(1-\cos^2\theta)
\nonumber
\\
&&
~~~~~~\times
{1\over \sinh \pi \xi_f}{1\over \sinh \pi \xi_i}{\kappa \over \sinh \pi \kappa}
{1\over (\xi_f-\xi_i+\kappa)\sinh \pi(\xi_f-\xi_i+\kappa)/2}
\nonumber
\\
&&~~~~~~\times
{\sinh \pi(\xi_f+\xi_i-\kappa)/2\over\xi_f+\xi_i-\kappa}
{\sinh \pi(\xi_f+\xi_i+\kappa)/2\over\xi_f+\xi_i+\kappa}
{\sinh \pi(\xi_f-\xi_i-\kappa)/2\over\xi_f-\xi_i-\kappa}.
\label{energyE}
\end{eqnarray}

\begin{figure*}[t]
\includegraphics[scale=0.5]{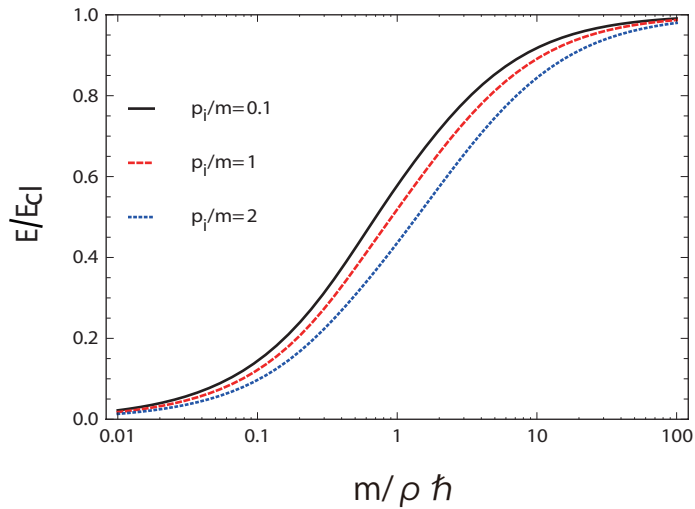}
\caption{The ratio of the total radiation energy to the 
classical WKB formula, $E/E_{\rm cl}$, as function of 
$m/\rho\hbar$ 
with fixing $p_i/m=0.1$ (black solid curve), $p_i/m=1$ (red dashed curve), and
$p_i/m=2$ (blue dotted curve). 
\label{fig2} {}}
\end{figure*}

\begin{figure*}[h]
\includegraphics[scale=0.5]{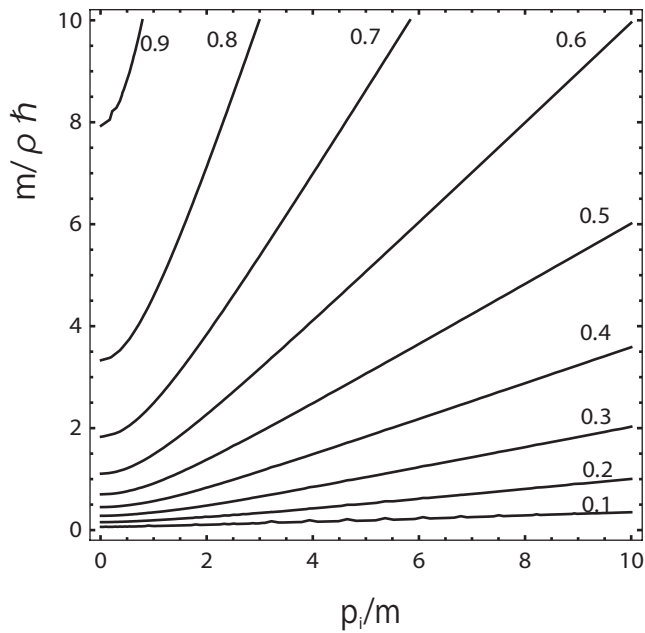}
\caption{The deviation from the classical WKB formula. 
Contour of $E/E_{\rm cl}$ on the $p_i/m$ and $m/\rho\hbar$ plane. 
\label{fig3} {}}
\end{figure*}

The classical radiation energy corresponding to the Larmor
radiation can be evaluated using Eq.~(\ref{EE0}) (see also 
Eq.~(36) of Ref.~\cite{NSY}).
In the case of the background universe with Eq.~(\ref{amilne}), we have
\begin{eqnarray}
E_{\rm cl}={e^2p_i^2\rho\over 24\pi\epsilon_0 m^2}.
\label{Eclex}
\end{eqnarray}
Combining (\ref{energyE}) and (\ref{Eclex}), we have
\begin{eqnarray}
{E\over E_{\rm cl}}&=&12\left(m\over \rho\hbar\right)^2
\int_0^\infty d\kappa \kappa^2 
\int_{-1}^1d\cos\theta(1-\cos^2\theta)
\nonumber
\\
&&~~~~~~\times
{1\over \sinh \pi \xi_f}{1\over \sinh \pi \xi_i}{\kappa \over \sinh \pi \kappa}
{1\over (\xi_f-\xi_i+\kappa)\sinh \pi(\xi_f-\xi_i+\kappa)/2}
\nonumber
\\
&&~~~~~~\times
{\sinh \pi(\xi_f+\xi_i-\kappa)/2\over\xi_f+\xi_i-\kappa}
{\sinh \pi(\xi_f+\xi_i+\kappa)/2\over\xi_f+\xi_i+\kappa}
{\sinh \pi(\xi_f-\xi_i-\kappa)/2\over\xi_f-\xi_i-\kappa}.
\label{EE}
\end{eqnarray}
We performed the integration of Eq.~(\ref{EE}) numerically.
Figure \ref{fig2} plots ${E/E_{\rm cl}}$ as a function of $m/\rho\hbar$
with fixing $p_i/m=0.1$ (black solid curve), 
$p_i/m=1$ (red dashed curve), and $p_i/m=2$ (blue dotted curve). 
In the limit $m/\rho\hbar\gg1$, one can see that 
${E/ E_{\rm cl}}$ approaches to $1$. 
On the other hand, in the limit  $m/\rho\hbar\ll1$, 
${E/ E_{\rm cl}}$ approaches to $0$.
Complimentary to this figure, Figure \ref{fig3} shows the 
contour of ${E/E_{\rm cl}}$ on the plane $p_i/m$ and $m/\rho\hbar$. 
For the region $m/\rho\hbar\gg1$, $E/E_{\rm cl}$ approaches to $1$, 
while for the region $m/\rho\hbar\ll1$, $E/E_{\rm cl}$ approaches to $0$.
These features are the same as those in the previous paper
\cite{NSY}, though a different background universe was
considered there. 

\section{WKB approach}
In the previous section, we investigated the radiation energy 
in the case when the mode function is exactly solved in the
analytic manner. 
From the result of the exactly solvable model in the 
previous section as well as the previous paper \cite{NSY}, 
the quantum correction reduces the total radiation energy compared with
the classical Larmor formula. 
In this section, we consider the origin of the 
quantum effect in the Larmor radiation. 
To this end, we adopt the WKB approach. 
Our computation is a generalization of the 
work by Higuchi and Walker \cite{HW}, in which 
the quantum effect of the Larmor radiation on an  
electric-field background is investigated (cf.~\cite{YNK}).

The mode function with the WKB approximation is
\begin{eqnarray}
\varphi_\bfp(\eta)={1\over \sqrt{2\Omega_\bfp}} \exp\left[
-i\int^\eta \Omega_\bfp(\eta') d\eta'
\right]
\label{WKBsolution}
\end{eqnarray}
with
\begin{eqnarray}
&&\Omega_\bfp(\eta)={\sqrt{\bfp^2+m^2a^2(\eta)}\over \hbar}.
\end{eqnarray}
Substituting the WKB solution (\ref{WKBsolution}) into 
Eq.~(\ref{radiationenergy3}),
we have
\begin{eqnarray}
E&=& -{e^2\over 2\epsilon_0}{1\over 2^2} \int {d^3k\over (2\pi)^3}   
\bigg\{
\biggl|\int d\eta {{\hat \bfk}\cdot(2\bfp_i-\hbar\bfk)
\over \hbar \sqrt{\Omega_{\bfp_i}}\sqrt{\Omega_{\bfp_f}}}
\exp\Bigl[ik\eta+i\int^\eta(\Omega_{\bfp_f}-\Omega_{\bfp_i})d\eta'\Bigr]
\biggr|^2
\nonumber
\\
&&~~~~~~~~~~~~-\biggl|
\int d\eta {2\bfp_i-\hbar\bfk
\over \hbar \sqrt{\Omega_{\bfp_i}}\sqrt{\Omega_{\bfp_f}}}
\exp\Bigl[ik\eta+i\int^\eta(\Omega_{\bfp_f}-\Omega_{\bfp_i})d\eta'\Bigr]
\biggr|^2
\bigg\}
\end{eqnarray}
with
\begin{eqnarray}
&&\Omega_{\bfp_i}(\eta)=\sqrt{\bfp_i^2+m^2a^2(\eta)}/\hbar,
\\
&&\Omega_{\bfp_f}(\eta)=\sqrt{(\bfp_i-\hbar\bfk)^2+m^2a^2(\eta)}/\hbar.
\end{eqnarray}
We derive the expression for the radiation energy 
in the form expanded as a power series with respect to $\hbar$.
%
Using the approximation,
\begin{eqnarray}
&&\Omega_{\bfp_f}-\Omega_{\bfp_i}
\simeq-{\bfk\cdot\bfp_i\over\sqrt{\bfp_i^2+m^2a^2(\eta)}}+{\hbar \over 2} \left(
{k^2\over\sqrt{\bfp_i^2+m^2a^2(\eta)} }-
{(\bfk\cdot\bfp_i)^2\over
\sqrt{\bfp_i^2+m^2a^2(\eta)}^3}\right),
\nonumber
\\
&&{1\over \hbar \sqrt{\Omega_{\bfp_i}}\sqrt{\Omega_{\bfp_f}}}
\simeq{1\over \sqrt{\bfp_i^2+m^2a^2(\eta)}}\bigg\{
1+{\hbar\over 2}{\bfk\cdot\bfp_i\over \bfp_i^2+m^2a^2(\eta)}
\biggr\},
\nonumber
\end{eqnarray}
we have the expression of the radiation energy up to the order of $\hbar$,
\begin{eqnarray}
E
&=&- 
{e^2\over 2\epsilon_0} \int {d^3k\over (2\pi)^3}
\int d\xi \int d\xi'
\left\{\Bigl({{\hat \bfk}\cdot{d\bfx\over d\xi}}\Bigr)
       \Bigl({{\hat \bfk}\cdot{d\bfx'\over d\xi'}}\Bigr)
-{d\bfx\over d\xi}\cdot{d\bfx'\over d\xi'}
\right\}e^{ik\xi-ik\xi'}
\nonumber
\\
&&\times\biggl[
1+{\hbar k\over 2}{\hat\bfk\cdot\bfp_i\over \bfp_i^2+m^2a^2(\eta)}
+{\hbar k\over 2}{\hat\bfk\cdot\bfp_i\over \bfp_i^2+m^2a^2(\eta')}
\nonumber
\\
&&~~~~+{i\hbar\over 2}\int_{\eta'}^\eta
\biggl({k^2\over \sqrt{\bfp_i^2+m^2a^2(\eta'')}}
-{k^2(\hat\bfk\cdot\bfp_i)^2\over \sqrt{\bfp_i^2+m^2a^2(\eta'')}^3}
\biggr)d\eta''
+{\cal O}(\hbar^2)
\biggr],
\label{Eexpression}
\end{eqnarray}
where we introduced the new variable $\xi$ instead of $\eta$
\begin{eqnarray}
\xi=\eta-\int^\eta {{\hat \bfk}\cdot\bfp_i
\over \sqrt{\bfp_i^2+m^2a^2(\eta')}}d\eta'.
\end{eqnarray}
Furthermore, we introduce the notation, 
\begin{eqnarray}
&&{d\bfx^i\over d\tau}=\bfp_i,
\label{dexxi}
\\
&&{d\eta\over d\tau}=\sqrt{\bfp_i^2+m^2a^2(\eta)}.
\label{dexeta}
\end{eqnarray}
Then, Eq.~(\ref{Eexpression}) is rephrased as
\begin{eqnarray}
E&=&-
{e^2\over 2\epsilon_0}\int {d^3k\over (2\pi)^3}
\int d\xi \int d\xi'
\left\{\Bigl({{\hat \bfk}\cdot{d\bfx\over d\xi}}\Bigr)
       \Bigl({{\hat \bfk}\cdot{d\bfx'\over d\xi'}}\Bigr)
-{d\bfx\over d\xi}\cdot{d\bfx'\over d\xi'}
\right\}e^{ik\xi-ik\xi'}
\nonumber
\\
&&\times\biggl[
1+{\hbar k\over 2}\biggl(
\hat\bfk\cdot{d\bfx\over d\eta} {d\tau\over d\eta}+
\hat\bfk\cdot{d\bfx'\over d\eta'} {d\tau'\over d\eta'}
\biggr)
+{i\hbar k^2\over 2}\int_{\tau({\xi'})}^{\tau(\xi)} d\tau''
\biggl(1-\biggl(\hat\bfk\cdot{d\bfx''\over d\eta''}\biggr)^2
\biggr)\biggr].
\label{EexpressionII}
\end{eqnarray}
We replace $k$ in Eq.~(\ref{EexpressionII}) with the partial 
derivative with respect to $\xi$ or $\xi'$  operating on $e^{ik\xi-ik\xi'}$,
which is the equivalent technique adopted in Ref.~\cite{HW}. 
Partial integrations lead to
\begin{eqnarray}
E^{(0)}&=&-
{e^2\over 2 \epsilon_0(2\pi)^3} \int d\hat\bfk \int_0^\infty dk  \int d\xi \int d\xi' 
e^{ik(\xi-\xi')}
\biggl(\Bigl({{\hat \bfk}\cdot{d^2\bfx\over d\xi^2}}\Bigr)
       \Bigl({{\hat \bfk}\cdot{d^2\bfx'\over d\xi'^2}}\Bigr)
-{d^2\bfx\over d\xi^2}\cdot{d^2\bfx'\over d\xi'^2}
\biggr),
\label{E0}
\\
E^{(1)}&=&-
{e^2\over 2 \epsilon_0(2\pi)^3} \int d\hat\bfk \int_0^\infty dk  \int d\xi \int d\xi' 
e^{ik(\xi-\xi')}
\nonumber
\\
&&\times\Biggl\{{i\hbar\over 4}\biggl({d\over d\xi}-{d\over d\xi'}\biggr)
{d\over d\xi}{d\over d\xi'}\biggl[\biggl(\Bigl({{\hat \bfk}\cdot{d\bfx\over d\xi}}\Bigr)
       \Bigl({{\hat \bfk}\cdot{d\bfx'\over d\xi'}}\Bigr)
-{d\bfx\over d\xi}\cdot{d\bfx'\over d\xi'}
\biggr)\biggl(
\hat\bfk\cdot{d\bfx\over d\eta} {d\tau\over d\eta}+
\hat\bfk\cdot{d\bfx'\over d\eta'} {d\tau'\over d\eta'}
\biggr)\biggr]
\nonumber
\\
&&~~~+
{i\hbar \over 2}
{d^2\over d\xi^2}{d^2\over d\xi'^2}\biggl[\biggl(\Bigl({{\hat \bfk}\cdot{d\bfx\over d\xi}}\Bigr)
       \Bigl({{\hat \bfk}\cdot{d\bfx'\over d\xi'}}\Bigr)
-{d\bfx\over d\xi}\cdot{d\bfx'\over d\xi'}
\biggr)
\int_{\xi'(\eta')}^{\xi(\eta)} d\xi''{d\tau''\over d\xi''}
\biggl(1-\biggl(\hat\bfk\cdot{d\bfx''\over d\eta''}\biggr)^2
\biggr)\biggr]\Biggl\},
\label{E1}
\end{eqnarray}
where $E^{(0)}$ and $E^{(1)}$ are the contribution of the order of 
 $\hbar^0$ and $\hbar$, respectively.
The expression of the order of $\hbar^0$, Eq.~(\ref{E0}), gives the 
classical radiation energy formula \cite{NSY,Higuchicfu} 
(cf.~\cite{HM,HMII,HMIII}), while the expression of the
order of $\hbar$, Eq.~(\ref{E1}), gives the first order 
quantum effect in the radiation \cite{HW,YNK}. 
We choose the coordinate $z$ to be parallel to 
${\bfp}_i$, and the polar angle $\theta$ to be the angle between 
the $z$-axis and photon momentum vector $\bfk$.
Then, we write
\begin{eqnarray}
&&v={dz\over d\eta},
\label{defvv} 
\\
&& {d\xi \over d\eta}=(1-v\cos\theta),
\label{defvvv} 
\\
&&{d^2z\over d\xi^2}={\dot v\over (1-v\cos\theta)^3} ,
\label{defvvvv} 
\end{eqnarray}
with which we have
\begin{eqnarray}
E^{(0)}&=&
{e^2\over 8\pi\epsilon_0}\int_{-1}^{1} d\cos\theta(1-\cos^2\theta)\int d\xi 
\left({d^2z\over d\xi^2}\right)^2
\label{E0b}
\end{eqnarray}
from Eq.~(\ref{E0}). Here, the `{\it dot}' denotes the 
differentiation with respect to $\eta$.
Integration with respect to $\theta$ yields
\begin{eqnarray}
E^{(0)}&=&
{e^2\over 6\pi\epsilon_0}\int d\eta {\dot v^2\over (1-v^2)^3}.
\label{E0d}
\end{eqnarray}
With the use of the relation $v=p_i/\sqrt{p_i^2+m^2a^2(\eta)}$,
Eq.~(\ref{E0d}) reduces to 
\begin{eqnarray}
&&E^{(0)}={e^2p_i^2\over 6\pi \epsilon_0m^2}\int d\eta {\dot a^2\over a^4}.
\label{EE0}
\end{eqnarray}
This is the classical radiation formula in an expanding universe which
can be regarded as the Larmor radiation \cite{NSY}.

Next, we consider the contribution of the order of $\hbar$,  Eq.~(\ref{E1}). 
Let us first consider the limit of the non-relativistic motion, 
$v\ll1$,
we can write the radiation energy in terms of the velocity and its time
derivatives, and find the leading contribution 
(see Eq.~(\ref{nonrelaapap}) in Appendix A),
\begin{eqnarray}
E^{(1)}&=&
{3e^2\hbar\over 4 \epsilon_0(2\pi)^2}
\int d\cos\theta\int d\eta \int d\eta' 
{1\over \eta-\eta'}\nonumber\\
&&\times 
\left[\frac{1-\cos^2\theta}{p_i}
\Bigl\{(\ddot{v}\dot{v}^\prime v^\prime-\ddot{v}^\prime\dot{v}v)+
\dot{v}\dot{v}^\prime\bigl(3(\dot{v}v^\prime-\dot{v}^\prime v)-
(\dot{v}v-\dot{v}^\prime v^\prime)\bigr)\cos\theta\Bigr\}\right],
\label{nonun}
\end{eqnarray}
where the `{\it dash}' denotes the function of $\eta'$, i.e., $v'=v(\eta')$.
Note that the latter terms of the right-hand-side of Eq.~(\ref{nonun}),
which is in proportion to $\cos\theta$, give no contribution 
when integrated over $\theta$. In the  non-relativistic case, we have 
\begin{eqnarray}
 \biggl|\frac{\ddot{v}}{\dot{v}^2}\biggr|\simeq\frac{ma}{p_i}
 \left|\frac{\ddot{a}a-2\dot{a}^2}{\dot{a}^2}\right|.
\end{eqnarray}
This gives a rough estimation of the ratio of the first term 
$(\ddot{v}\dot{v}^\prime v^\prime-\ddot{v}^\prime\dot{v}v)$ in
the right-hand-side of Eq.~(\ref{nonun}) to the 
latter terms in proportion to $\cos\theta$. 
Therefore, the first term is larger than the latter terms
in the  non-relativistic case, $ma/p_i\gg1$.

Next, we consider the relativistic limit, $v\sim 1$. 
In an expanding universe, the condition of the relativistic motion
is guaranteed as long as $ma(\eta)\simlt p_i$. Hence, the particle
motion becomes non-relativistic as the universe expands with (\ref{amilne}), 
even though the motion is relativistic initially. In the model with 
(\ref{amilne}), the particle motion is relativistic for
\begin{eqnarray}\label{milnecond}
{m\over p_i}e^{\rho\eta}\ll 1.
\end{eqnarray}
On the other hand, the ratio of $\ddot{v}$ to $\dot{v}^2$ is written as
\begin{eqnarray}\label{relamilne}
  \left|\frac{\ddot{v}}{\dot{v}^2}\right|\simeq
\frac{p_i^2}{m^2}\left|\frac{\ddot{a}a+\dot{a}^2}{(\dot{a}a)^2}\right|
\simeq \frac{p_i^2}{m^2}e^{-2\rho\eta}.
\end{eqnarray}
This means that $|\dot{v}^2|\ll|\ddot{v}|$
while the particle motion is relativistic. 
In this case, 
using expression (\ref{relatisticlimitap}) in Appendix A, we have
\begin{eqnarray}
E^{(1)}&\simeq&
{3e^2\hbar \over 4 \epsilon_0(2\pi)^2}
\int d\cos\theta\int d\eta \int d\eta' 
{1\over \xi(\eta)-\xi(\eta')}
\nonumber
\\
&&\times\left[
{1-\cos^2\theta\over p_i}
\left\{
{\ddot v\dot v'\over (1-v\cos\theta)^3(1-v'\cos\theta)^2}
-{\ddot v'\dot v\over (1-v\cos\theta)^2(1-v'\cos\theta)^3}
\right\}\right].
\end{eqnarray}
This expression is derived assuming the particle motion is relativistic 
all the time. Then, it is not justified when the particle motion
changes from relativistic to non-relativistic as the universe 
expands, as for the model with (\ref{amilne}) in Sec.~3.

\section{Discussions}
Integrating the right-hand-side of 
Eq.~(\ref{nonun}) with respect to $\theta$, we obtain the 
quantum correction to the Larmor 
radiation of the order of $\hbar$,
\begin{eqnarray}
&&E^{(1)}={e^2\hbar \over 4\pi^2 \epsilon_0p_i}
\int d\eta \int d\eta'{\ddot v\dot v' v'-\ddot v'\dot v v\over \eta-\eta'}.
\label{EE1a}
\end{eqnarray}
This result should be compared with the result by Higuchi and 
Walker \cite{HW}, which presented the quantum correction to the 
Larmor radiation of the order of $\hbar$ from a charged particle
in an accelerated motion on a homogeneous electric field background,
\begin{eqnarray}
&&E_{HW}^{(1)}={e^2\hbar \over 6\pi^2 \epsilon_0p_i}
\int dt \int dt'{\ddot v\dot v'-\ddot v'\dot v \over t-t'}.
\label{EE1aHW}
\end{eqnarray}
There is a difference between (\ref{EE1a}) and (\ref{EE1aHW}), i.e.,
the numerical coefficient and the dependence on the velocity and its 
derivatives. However,  the quantum effect is non-local in time in both cases.


In the case when the velocity $v$ and its derivatives are approximated as
\begin{eqnarray}
&&v={p_i\over \sqrt{p_i^2+m^2 a^2(\eta)}}\simeq{p_i\over m a(\eta)},
\\
&&\dot v\simeq-{p_i\over m}{\dot a\over  a^2},
\label{nonra}
\\
&&\ddot v\simeq-{p_i\over m}{\ddot a a-2\dot a^2 \over a^3},
\label{nonrb}
\end{eqnarray}
Eq.~(\ref{EE1a}) yields
\begin{eqnarray}
&&E^{(1)}={e^2\hbar \over 4\pi^2\epsilon_0 m}{p_i^2\over m^2}\int d\eta \int d\eta' {1\over \eta-\eta'}
\left({(\ddot a a-2\dot a^2)\dot a'-(\ddot a'a'-2\dot a'^2)\dot a \over a^3 a'^3}\right),
\label{EEb}
\end{eqnarray}
where $a=a(\eta)$ and $a'=a(\eta')$.

\begin{figure*}[t]
\includegraphics[scale=0.5]{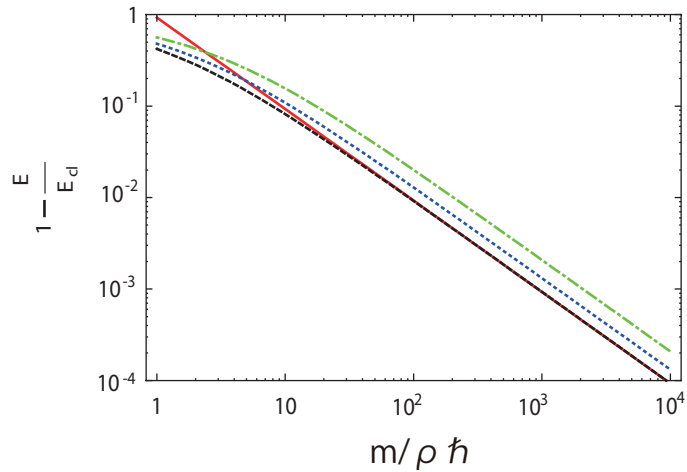}
\caption{Comparison of the deviation of the classical 
radiation from the total radiation. The solid (red) 
curve is $-E^{(1)}/E_{\rm cl}$ as a function of $m/\rho\hbar$, 
while the other curves are $-(E-E_{\rm cl})/E_{\rm cl}$ with fixing 
$p_i/m=0.01$ (black dashed curve), $p_i/m=1$ (blue dotted curve), 
and $p_i/m=2$ (green dash-dotted curve), where 
$E^{(1)}$ is computed using Eq.~(\ref{EEb}).
\label{fig4} {}}
\end{figure*}

Figure \ref{fig4} shows the deviation of the total Larmor 
radiation from the classical Larmor radiation from a moving 
charge in the universe considered in Sec.~3, as a function of $m/\rho\hbar$.
The curves compare the case when the radiation is computed
with the exact mode function of Sec.~3 and the case
when the radiation is computed with the WKB approaches up 
to the order of $\hbar$.
The (black) dashed curve, the (blue) dotted curve, and the 
(green) dash-dotted curve are the results with the exact mode 
function with $p_i/m=0.01,~1$, and $2$, respectively.
The solid (red) curve is from the WKB mode function, Eq.~(\ref{EEb}).
For the region with $p_i/m\ll 1$ and $m/\rho\hbar \gg 1$, the quantum 
correction can be well explained by the WKB approximate formula,  
Eq.~(\ref{EEb}).
The condition $p_i/m\ll 1$ comes from the fact that Eq.~(\ref{EEb}) 
is derived in the non-relativistic approximation. 
The deviation of the results with $p_i/m=1$ and $2$ from Eq.~(\ref{EEb}) 
comes from the fact that the particle motion is relativistic initially.
However, the deviation is not significant, which may indicate that
the fraction of the radiation from the particle in the non-relativistic 
regime is not small. 

Finally, we show that the condition $m/\rho\hbar \gg 1$ is the 
necessary condition for the WKB approximation in an explicit 
manner. The WKB approximation holds under the condition \cite{BD},  
\begin{eqnarray}
 \frac{1}{2\Omega^2}
\left|\frac{\ddot{\Omega}}{\Omega}-\frac{3}{2}
\frac{\dot{\Omega}^2}{\Omega^2}\right|
\ll1. 
\label{WKB}
\end{eqnarray}
In the universe with the scale factor 
(\ref{amilne}), the condition (\ref{WKB}) is
written as 
\begin{eqnarray}
\left({\hbar \rho \over m}\right)^2
\biggl|{(a^2-1)(a^2-5-4p_i^2/m^2)\over 4(a^2+p_i^2/m^2)^3}\biggr|
\ll1.
\label{WKB2}
\end{eqnarray}
Hence, the use of the WKB approximation is guaranteed 
when $m/\rho\hbar \gg 1$ is satisfied. 
This condition is interpreted as follows \cite{NSY}. 
In the universe with (\ref{amilne}), 
the ratio of the Compton wavelength to the Hubble horizon is 
\begin{eqnarray}
  {\hbar \rho \over m} {a^2-1\over a^3}. 
\label{WKB3}
\end{eqnarray}
When the condition $m/\rho\hbar \gg 1$ is satisfied, 
the ratio (\ref{WKB3}) is much less than unity. Thus, the
condition for the WKB approximation can be regarded as
the condition that the Compton wavelength is much shorter
than the Hubble horizon. 

\section{conclusions}
In this paper, we investigated the quantum effect of the Larmor 
radiation from a moving charge in an expanding universe based 
on the framework of the perturbative approach of the scalar QED theory.
We first derived the general formula for the quantum radiation energy, 
which is expressed by the mode function of the free complex scalar
field. Assuming the background universe that the Minkowski spacetime
transits to the Milne universe, in which the mode function 
is exactly solved in the analytic manner, we demonstrated
how the total radiation energy deviates from the classical Larmor
radiation formula. This gives us a hint to understand the quantum 
effect on the Larmor radiation. 

In the latter part of the paper, we investigated the quantum 
effect of the Larmor radiation with the use of the WKB approximation
in constructing the mode function.
In the WKB approach, 
which is valid as long as the Compton wavelength is shorter than the 
Hubble horizon length, 
we investigated the quantum effect of the 
Larmor radiation by evaluating the contribution of the order of 
$\hbar$ of the radiation energy with the approximate mode 
function. The radiation of the order of $\hbar$ well explains
the deviation of the total radiation with the exact approach 
from the classical Larmor formula in Sec.~3.
The radiation of the order of $\hbar$ is determined by 
the non-local integration in time depending the behavior 
of the expansion of the background universe, as in the case 
of a homogeneous electric field background \cite{HW,YNK}.
However, the first order quantum correction to the Larmor 
radiation in the conformally flat universe is not the 
same as that on the homogeneous electric field background. 

\vspace{0.2cm}
{\it Acknowledgements}
K.Y. thanks A.~Higuchi, H. Nomura and M. Sasaki for useful communication
when he initiated the study of the present paper. He also thanks Y. Nambu
for useful discussions. We are grateful for anonymous referee for useful
comments. This work was supported by Japan Society for Promotion of 
Science (JSPS) Grants-in-Aid for Scientific Research 
(Nos.~21540270,~21244033). This work is also supported by JSPS 
Core-to-Core Program ``International Research 
Network for Dark Energy''.
All the numerical computation presented in this paper were 
performed with the help of the package MATHEMATICA version 7.

%
%


\appendix
\section{useful formulas}
\label{183921_15Nov10}
In this Appendix, we summarize useful formulas in deriving the 
radiation formula in the present paper. 
Using Eqs.~(\ref{dexxi}), (\ref{dexeta}), and  
(\ref{defvv})-(\ref{defvvvv}), we have 
\begin{eqnarray}
&&{d^3\bfz\over d\xi^3}
={\ddot v\over (1-v\cos\theta)^4}+{3\dot v^2\cos\theta\over (1-v\cos\theta)^5},
\nonumber
\end{eqnarray}
\begin{eqnarray}
&&{d\tau\over d\eta}={1\over \sqrt{\bfp_i^2+m^2a^2(\eta)}},
~~~~
{d\over d\xi}\left({v{d\tau\over d\eta}}\right)=
{2\dot v v\over p_i(1-v\cos\theta)},
~~~~
{d^2\over d\xi^2}\left({v{d\tau\over d\eta}}\right)=
{2\dot v^2\over p_i(1-v\cos\theta)^3}+{2\ddot v v\over p_i(1-v\cos\theta)^2},
\nonumber
\end{eqnarray}
\begin{eqnarray}
&&{d\over d\xi}\left({v{d\tau\over d\eta}}\right)={1\over 1-v\cos\theta}\left(
{\dot v\over\sqrt{\bfp_i^2+m^2a^2(\eta)}}-{ v m^2 a(\eta)\dot a(\eta)\over\sqrt{\bfp_i^2+m^2a^2(\eta)}^3}
\right),
\nonumber
\\
&&{d^2\over d\xi^2}\left({v{d\tau\over d\eta}}\right)={\dot v\cos\theta
\over (1-v\cos\theta)^3}\left(
{\dot v\over\sqrt{\bfp_i^2+m^2a^2(\eta)}}-{v m^2 a(\eta)\dot a(\eta)\over\sqrt{\bfp_i^2+m^2a^2(\eta)}^3}
\right)
+{1\over (1-v\cos\theta)^2}\Biggl(
{\ddot v\over\sqrt{\bfp_i^2+m^2a^2(\eta)}}
\nonumber
\\
&&~~~~~~~~~~~~~~~~
-{2\dot v m^2 a(\eta)\dot a(\eta)+v m^2\dot a(\eta)\dot a(\eta)
+v m^2 a(\eta)\ddot a(\eta) \over\sqrt{\bfp_i^2+m^2a^2(\eta)}^3}
+{3v (m^2 a(\eta)\dot a(\eta))^2\over\sqrt{\bfp_i^2+m^2a^2(\eta)}^5}\Biggr),
\nonumber
\end{eqnarray}
where the dot denotes the differentiation with respect to the
conformal time $\eta$.
Using these relations and 
$\hat \bfk\cdot d\bfx/d\eta=\cos\theta dz/d\eta=v\cos\theta $,  
it is straightforward to derive
\begin{eqnarray}
&&\hspace{-1cm}\biggl({d\over d\xi}-{d\over d\xi'}\biggr)
{d\over d\xi}{d\over d\xi'}\biggl[\biggl(\Bigl({{\hat \bfk}\cdot{d\bfx\over d\xi}}\Bigr)
       \Bigl({{\hat \bfk}\cdot{d\bfx'\over d\xi'}}\Bigr)
-{d\bfx\over d\xi}\cdot{d\bfx'\over d\xi'}
\biggr)\biggl(
\hat\bfk\cdot{d\bfx\over d\eta} {d\tau\over d\eta}+
\hat\bfk\cdot{d\bfx'\over d\eta'} {d\tau'\over d\eta'}
\biggr)\biggr]
\nonumber\\
&=&(\cos^2\theta-1)\cos\theta\Biggl\{\biggl[       
\biggl({\ddot v\over (1-v\cos\theta)^4}+{3\dot v^2\cos\theta\over (1-v\cos\theta)^5}\biggr)
\biggl({\dot v'\over (1-v'\cos\theta)^3}\biggr)
\nonumber
\\
&&~~~~~~~~~~~~~~~~~~~~~-
\biggl({\dot v\over (1-v\cos\theta)^3}\biggr)
\biggl({\ddot v'\over (1-v'\cos\theta)^4}+{3\dot v'^2\cos\theta\over (1-v'\cos\theta)^5}\biggr)\biggr]
 \biggl({v^2+v'^2\over p_i}\biggr)
\nonumber\\
&&+2
\biggl({\dot v\over (1-v\cos\theta)^3}\biggr)
\biggl({\dot v'\over (1-v'\cos\theta)^3}\biggr)
   \biggl({2v\dot v\over p_i(1-v\cos\theta)}-{2v'\dot v'\over p_i(1-v'\cos\theta)}
  \biggr)
\nonumber\\
&&+
\biggl({v\over 1-v\cos\theta}\biggr)
\biggl({\dot v'\over (1-v'\cos\theta)^3}\biggr)
\biggl({2\dot v^2\over (1-v\cos\theta)^3}+{2v\ddot v\over (1-v\cos\theta)^2}\biggr){1\over p_i}
\nonumber\\
&&-
\biggl({\dot v\over (1-v\cos\theta)^3}\biggr)
\biggl({v'\over 1-v'\cos\theta}\biggr)
\biggl({2\dot v'^2\over (1-v'\cos\theta)^3}+{2v'\ddot v'\over (1-v'\cos\theta)^2}\biggr){1\over p_i}
\nonumber\\
&&+
\biggl({\ddot v\over (1-v\cos\theta)^4}+{3\dot v^2\cos\theta\over (1-v\cos\theta)^5}\biggr)
\biggl({v'\over (1-v'\cos\theta)}\biggr)
{2v' \dot v'\over p_i(1-v'\cos\theta)}
\nonumber
\\
&&-
\biggl({v\over (1-v\cos\theta)}\biggr)
\biggl({\ddot v'\over (1-v'\cos\theta)^4}+{3\dot v'^2\cos\theta\over (1-v'\cos\theta)^5}\biggr)
{2v \dot v\over p_i(1-v\cos\theta)}\Biggr\},
\label{184219_15Nov10}
\end{eqnarray}
and
\begin{eqnarray}
&&\hspace{-1cm}{d^2\over d\xi^2}{d^2\over d\xi'^2}\biggl[\biggl(\Bigl({{\hat \bfk}\cdot{d\bfx\over d\xi}}\Bigr)
       \Bigl({{\hat \bfk}\cdot{d\bfx'\over d\xi'}}\Bigr)
-{d\bfx\over d\xi}\cdot{d\bfx'\over d\xi'}
\biggr)
\int_{\xi'(\eta')}^{\xi(\eta)} d\xi''{d\tau''\over d\xi''}
\biggl(1-\biggl(\hat\bfk\cdot{d\bfx''\over d\eta''}\biggr)^2
\biggr)\biggr]
\nonumber\\
&=&(\cos^2\theta-1)\Biggl\{\biggl(
 {d^3\bfz\over d\xi^3}{d^3\bfz'\over d\xi'^3}\biggr)
   \int_{\xi'}^\xi d\xi'' {d\tau''\over d\xi''}
   \biggl(1-\cos^2\theta\Bigl({d\bfz''\over d\eta''}\Bigr)^2\biggr)
\nonumber\\
&-&2
\biggl({\ddot v\over (1-v\cos\theta)^4}+{3\dot v^2\cos\theta\over (1-v\cos\theta)^5}\biggr)
\biggl({\dot v'\over (1-v'\cos\theta)^3}\biggr)
{v' (1+v'\cos\theta)\over p_i}
\nonumber\\
&+&2
\biggl({\dot v\over (1-v\cos\theta)^3}\biggr)
\biggl({\ddot v'\over (1-v'\cos\theta)^4}+{3\dot v'^2\cos\theta\over (1-v'\cos\theta)^5}\biggr)
{v (1+v\cos\theta)\over p_i}
\nonumber\\
&-&
\biggl({\ddot v\over (1-v\cos\theta)^4}+{3\dot v^2\cos\theta\over (1-v\cos\theta)^5}\biggr)
\biggl({v'\over 1-v'\cos\theta}\biggr)
{\dot v' (1+2v'\cos\theta)\over p_i(1-v'\cos\theta)}
\nonumber\\
&+&
\biggl({v\over 1-v\cos\theta}\biggr)
\biggl({\ddot v'\over (1-v'\cos\theta)^4}+{3\dot v'^2\cos\theta\over (1-v'\cos\theta)^5}\biggr)
{\dot v (1+2v\cos\theta)\over p_i(1-v\cos\theta)}\Biggr\}.
\label{184147_15Nov10}
\end{eqnarray}

In the limit of the non-relativistic motion, $v\ll1$, the above 
formulas give 
\begin{eqnarray}
&&{i\hbar\over 4}
\Biggl\{\biggl({d\over d\xi}-{d\over d\xi'}\biggr)
{d\over d\xi}{d\over d\xi'}\biggl[\biggl(\Bigl({{\hat \bfk}\cdot{d\bfx\over d\xi}}\Bigr)
       \Bigl({{\hat \bfk}\cdot{d\bfx'\over d\xi'}}\Bigr)
-{d\bfx\over d\xi}\cdot{d\bfx'\over d\xi'}
\biggr)\biggl(
\hat\bfk\cdot{d\bfx\over d\eta} {d\tau\over d\eta}+
\hat\bfk\cdot{d\bfx'\over d\eta'} {d\tau'\over d\eta'}
\biggr)\biggr]
\nonumber
\\
&&~~+
2
{d^2\over d\xi^2}{d^2\over d\xi'^2}\biggl[\biggl(\Bigl({{\hat \bfk}\cdot{d\bfx\over d\xi}}\Bigr)
       \Bigl({{\hat \bfk}\cdot{d\bfx'\over d\xi'}}\Bigr)
-{d\bfx\over d\xi}\cdot{d\bfx'\over d\xi'}
\biggr)
\int_{\xi'(\eta')}^{\xi(\eta)} d\xi''{d\tau''\over d\xi''}
\biggl(1-\biggl(\hat\bfk\cdot{d\bfx''\over d\eta''}\biggr)^2
\biggr)\biggr]\Biggl\}
\nonumber
\\
&&\simeq \frac{3i\hbar}{2}\left[\frac{1-\cos^2\theta}{p_i}
\left\{\ddot{v}\dot{v}^\prime v^\prime-\ddot{v}^\prime\dot{v}v+
\dot{v}\dot{v}^\prime\left(3(\dot{v}v^\prime-\dot{v}^\prime v)-
(\dot{v}v-\dot{v}^\prime v^\prime)\right)\cos\theta\right\}\right].
\label{nonrelaapap}
\end{eqnarray}
Here, we omitted the contribution from the term,
\begin{eqnarray}
2(\cos^2\theta-1)\biggl(
 {d^3\bfz\over d\xi^3}{d^3\bfz'\over d\xi'^3}\biggr)
   \int_{\xi'}^\xi d\xi'' {d\tau''\over d\xi''}
   \biggl(1-\cos^2\theta\Bigl({d\bfz''\over d\eta''}\Bigr)^2\biggr),
\label{integralterm}
\end{eqnarray}
because its leading contribution to the radiation energy in Eq.~(\ref{E1}) is
zero when $\dot v$ is zero at the boundary $\eta\rightarrow\pm\infty$. 

In the limit of the relativistic motion, $v\sim v'\sim1$, we have
\begin{eqnarray}
 &&{i\hbar\over 4}
\Biggl\{\biggl({d\over d\xi}-{d\over d\xi'}\biggr)
{d\over d\xi}{d\over d\xi'}\biggl[\biggl(\Bigl({{\hat \bfk}\cdot{d\bfx\over d\xi}}\Bigr)
       \Bigl({{\hat \bfk}\cdot{d\bfx'\over d\xi'}}\Bigr)
-{d\bfx\over d\xi}\cdot{d\bfx'\over d\xi'}
\biggr)\biggl(
\hat\bfk\cdot{d\bfx\over d\eta} {d\tau\over d\eta}+
\hat\bfk\cdot{d\bfx'\over d\eta'} {d\tau'\over d\eta'}
\biggr)\biggr]
\nonumber
\\
&&~~+
2
{d^2\over d\xi^2}{d^2\over d\xi'^2}\biggl[\biggl(\Bigl({{\hat \bfk}\cdot{d\bfx\over d\xi}}\Bigr)
       \Bigl({{\hat \bfk}\cdot{d\bfx'\over d\xi'}}\Bigr)
-{d\bfx\over d\xi}\cdot{d\bfx'\over d\xi'}
\biggr)
\int_{\xi'(\eta')}^{\xi(\eta)} d\xi''{d\tau''\over d\xi''}
\biggl(1-\biggl(\hat\bfk\cdot{d\bfx''\over d\eta''}\biggr)^2
\biggr)\biggr]\Biggl\}
\nonumber
\\
&&\simeq\frac{i\hbar}{4}\frac{6(1-\cos^2\theta)}{p_i(1-v\cos\theta)^2(1-v^\prime\cos\theta)^2}
\left\{{\ddot v\dot v'\over (1-v\cos\theta)^2(1-v'\cos\theta)}
-{\ddot v'\dot v\over (1-v\cos\theta)(1-v'\cos\theta)^2}
\right\}.
\label{relatisticlimitap}
\end{eqnarray}
Here, again we omitted the contribution from the term (\ref{integralterm})
because of the similar reason to that in the non-relativistic limit,
as long as the particle motion is relativistic all the time.
However, the omission is not justified when the particle motion
changes from relativistic to non-relativistic as the universe expands.


\begin{thebibliography}{99}
\bibitem{Iso}S. Iso, Y. Yamamoto, and S. Zhang, arXiv1011.4191
\bibitem{Homma} K. Homma, D. Habs, and T. Tajima, arXiv:1006.4533.
\bibitem{Gies} H. Gies, Eur. Phys. J. D{\bf 55}, (2009) 311.
\bibitem{Schwinger} J. Schwinger, Phys. Rev. {\bf 82}, 664 (1951)
\bibitem{Unruh} W. G. Unruh, Phys. Rev. D {\bf 14}, 870 (1976)
\bibitem{HiguchiPR} L. C. B. Crispino, A. Higuchi, and G. E. A. Matsas,
Rev.~Mod.~Phys.~{\bf 80}, 787 (2008) 
\bibitem{ChenTajima} P. Chen and T. Tajima, Phys. Rev. Lett. 
{\bf 83}, 256 (1999)
\bibitem{Schutzhold} R. Schutzhold, G. Schaller, and D. Habs, 
Phys. Rev. Lett. {\bf 97}, 121302 (2006)
\bibitem{YNK} K. Yamamoto and G. Nakamura, Phys. Rev. D, in press, 
arXiv:1012.5182
\bibitem{NSY} H. Nomura, M. Sasaki, and K. Yamamoto JCAP {\bf 0611} 013 (2006)
\bibitem{Higuchicfu} A. Higuchi and P. J. Walker, Phys. Rev. D {\bf 79}
105023 (2009) 
\bibitem{futamase} T. Futamase, M. Hotta, H. Inoue, and M. Yamaguchi, 
 Prog. Theor. Phys. {\bf 96}, 113 (1996)
\bibitem{HW}A. Higuchi and P. J. Walker, Phys. Rev. D {\bf 80} 105019 (2009)
\bibitem{Weinberg} S. Weinberg Phys. Rev. D {\bf 72} 043514 (2005)
\bibitem{Gradshteyn} I. S. Gradshteyn and I. M. Ryzhik, {\it Table of
Integrals, Series, and Products} (Academic Press, 1994)
\bibitem{HM} A. Higuchi and G. D. R. Martin, Found. Phys. {\bf 35} 1149 (2005)
\bibitem{HMII}
A. Higuchi and G. D. R. Martin, Phys. Rev. D {\bf 73} 025019 (2006)
\bibitem{HMIII}
A. Higuchi and G. D. R. Martin, Phys. Rev. D {\bf 74} 125002 (2006)
\bibitem{BD}
 N. D. Birrell and P. C. W. Davies, {\it Quantum fields in curved space} (Cambridge University Press ,1982)
\end{thebibliography}
\end{document}